\documentclass[aps,showpacs,twocolumn, prb]{revtex4-1}
\usepackage{hyperref}
\usepackage{graphicx}
\usepackage{amsmath,amssymb,amsfonts}
\usepackage{upgreek}
\usepackage{booktabs}
\usepackage{indentfirst}

\begin{document}

\title{Superconductivity in the Nb-Ru-Ge $\sigma$-Phase}

\author{Elizabeth M. Carnicom$^{1,*}$}
\author{Weiwei Xie$^2$}
\author{Zuzanna Sobczak$^3$}
\author{Tai Kong$^1$}
\author{Tomasz Klimczuk$^3$}
\author{Robert J. Cava$^{1,*}$}

\affiliation{$^1$Department of Chemistry, Princeton University, Princeton, New Jersey 08544}
\affiliation{$^2$Department of Chemistry, Louisiana State University, Baton Rouge, LA 70803}
\affiliation{$^3$Department of Physics, Gdansk University of Technology, Gdansk Poland 80-233}

\begin{abstract}

We show that the previously unreported ternary $\sigma$-phase material Nb$_{20.4}$Ru$_{5.7}$Ge$_{3.9}$ is a superconductor with a critical temperature of 2.2 K. Temperature-dependent magnetic susceptibility, resistance, and specific heat measurements were used to characterize the superconducting transition. The Sommerfeld constant $\gamma$ for Nb$_{20.4}$Ru$_{5.7}$Ge$_{3.9}$ is 91 mJ mol-f.u.$^{-1}$K$^{-2}$ and the specific heat anomaly at the superconducting transition, $\Delta$C/$\gamma$\textit{T}$_c$, is approximately 1.38. The zero-temperature upper critical field ($\mu_0$\textit{H}$_{c2}$(0)) was estimated to be 2 T by resistance data. Field-dependent magnetization data analysis estimated $\mu_0$\textit{H}$_{c1}$(0) to be 5.5 mT. Thus, the characterization shows Nb$_{20.4}$Ru$_{5.7}$Ge$_{3.9}$ to be a type II BCS superconductor.  This material appears to be the first reported ternary phase in the Nb-Ru-Ge system, and the fact that there are no previously reported binary Nb-Ru, Nb-Ge, or Ru-Ge $\sigma$-phases shows that all three elements are necessary to stabilize the material. A $\sigma$-phase in the Ta-Ru-Ge system was synthesized but did not display superconductivity above 1.7 K, which suggests that electron count cannot govern the superconductivity observed. Preliminary characterization of a possible superconducting $\sigma$-phase in the Nb-Ru-Ga system is also reported.  

\end{abstract}
\maketitle

\section{Introduction}

The sigma ($\sigma$) phases, which are typically brittle, have been extensively studied in materials science due to their detrimental effects on the mechanical properties of various steels, although the precipitation of this phase in specific amounts can sometimes lead to hardening as well.\cite{TheSigmaPhase,SigmaRev,Steel4,SteelRev,SteelMag} $\sigma$-phases, with the CrFe structure type and 30 atoms per unit cell, have extremely broad compositional existence ranges with complex compositions as a common feature - extensive substitutions on one or more of the five crystallographically distinct sites (2$a$, 4$f$, 8$i_1$, 8$i_2$ and 8$j$) in the structure have been reported.\cite{TheSigmaPhase} $\sigma$-phases are known to exist in over 40 different binary systems, and superconductivity has been observed in several of these.\cite{SigmaRev,Roberts} Nb$_{65.2}$Rh$_{34.8}$ for example, has been reported to display superconductivity with a critical temperature (\textit{T}$_c$) of 2.9 K\cite{NbRh_TaRh} and Nb$_{62}$Pt$_{38}$ has a \textit{T}$_c$ of 2.1 K.\cite{Nb_Pt1} Differing \textit{T}$_c$ values have been observed based on the composition of the $\sigma$-phase in both the Nb-Ir and Mo-Re binary systems.\cite{RT_MoRe,MoRe_SC} The binary $\sigma$-phase Mo-67$\%$ Re displays a \textit{T}$_c$ of 5.8 K, while Mo-50$\%$ Re has a \textit{T}$_c$ of 6.4 K.\cite{MoRe_SC,IBM} For the Nb-Ir system, the literature values of \textit{T}$_c$ vary from 2 K - 9 K for the $\sigma$-phases.\cite{Tc_sigmaphase,ConfProc,JChemSolids,Nb_Ir_SC1,Nb_Al1} The W-Os $\sigma$-phase shows similar behavior, where \textit{T}$_c$ varies from 2.5 K - 3.8 K as the osmium content is increased.\cite{IBM} The changes in \textit{T}$_c$ based on composition for the $\sigma$-phases is consistent with arguments that the critical temperature increases as the unit cell volume decreases and the valence electron count (VEC) per atom increases.\cite{IBM, JChemSolids}

	Here we report the new superconducting $\sigma$-phase Nb$_{20.4}$Ru$_{5.7}$Ge$_{3.9}$. To the best of our knowledge, this is the first ternary phase in the Nb-Ru-Ge system. Its superconducting transition is sharp and reproducible from one preparation to the next, and powder XRD patterns show the appearance of second phases when deviations from this composition are made, and thus the material forms in a relatively narrow composition range. In addition, a $\sigma$-phase has not been reported in the Ge-Ru, Nb-Ge, or Nb-Ru binary systems, showing that all three elements are necessary to stabilize the $\sigma$-phase material. We present the crystal structure determined by single crystal X-ray diffraction and characterize the superconducting transition through temperature-dependent magnetic susceptibility, resistance, and specific heat measurements. All measurements consistently show a critical temperature of 2.2 K. Specific heat data confirms that the transition is from the bulk of the material and Nb$_{20.4}$Ru$_{5.7}$Ge$_{3.9}$ appears to be a weak coupled BCS-type superconductor.  A Ta-Ru-Ge $\sigma$-phase, with an approximate composition of Ta$_{20.4}$Ru$_{5.7}$Ge$_{3.9}$, was synthesized but did not display superconductivity above 1.7 K. We also present preliminary results of the possible 2 K superconductor in the Nb-Ru-Ga $\sigma$-phase system.
	 
\section{Experimental Methods}
 
The starting materials for the synthesis of polycrystalline Nb$_{20.4}$Ru$_{5.7}$Ge$_{3.9}$ (or the tantalum or gallium variant) were niobium ($>$99.9$\%$, 325 mesh, Aldrich), tantalum ($>$99.9$\%$, foil, 0.127mm, Alfa), ruthenium ($>$99.9$\%$, 200 mesh, Aldrich), germanium ($>$99.9$\%$, 3.2mm, Alfa), and gallium ($>$99.99$\%$, pellet, 6mm dia., Aldrich). Niobium and ruthenium powders were pressed into pellets and first arc-melted separately in order to avoid significant mass loss during melting with germanium. The niobium (or tantalum), ruthenium, and germanium chunks were then arc-melted in a Zr-gettered atmosphere under $\sim$600 mbar Ar in a 6.8:1.9:1.3 (20.4:5.7:3.9) ratio. The purest sample for Nb-Ru-Ga system resulted when the loading composition was Nb$_{20}$Ru$_5$Ga$_5$. In addition, variation of the composition from the above formulas led to the presence of second phases in significant amounts. The arc-melted button was flipped over and remelted 3 times in order to ensure homogeneity throughout the sample. Mass loss after melting was $<$1$\%$. Samples of Nb$_{20.4}$Ru$_{5.7}$Ge$_{3.9}$ are stable and do not decompose over time when exposed to air. Room temperature powder X-ray diffraction (pXRD) was used to determine the purity of the samples using a Bruker D8 Advance Eco Cu K$_\alpha$ radiation ($\lambda$ =1.5406 $\textsc{\AA}$) diffractometer equipped with a LynxEye-XE detector. Annealing at temperatures below 1200 $^{\circ}$C did not yield more pure materials. Single crystals taken from the as-melted sample were mounted on the tips of Kapton loops and room temperature intensity data were collected using a Bruker Apex II X-ray diffractometer with Mo K$_{\alpha1}$ radiation ($\lambda$=0.71073 $\textsc{\AA}$). All data were collected over a full sphere of reciprocal space with 0.5$^{\circ}$ scans in $\omega$ and an exposure time of 10 s per frame. The 2$\theta$ range was from 4$^{\circ}$ to 75$^{\circ}$ and the SMART software was used for acquiring all data. The SAINT program was used to extract intensities and correct for Lorentz and polarization effects. Numerical absorption corrections were done with XPREP, which is based on face-indexed absorption.\cite{SHELXTL} For the single crystal refinement, the formula was constrained to the composition Nb$_{20.4}$Ru$_{5.7}$Ge$_{3.9}$ since this loading composition resulted in a single phase sample and the arc-melting process had $<$1$\%$ mass loss. The crystal structure of Nb$_{20.4}$Ru$_{5.7}$Ge$_{3.9}$ was solved using direct methods and refined by full-matrix least-squares on F$^2$ using the SHELXTL package.\cite{SHELX} All crystal structure drawings were created in the program VESTA.\cite{VESTA} A Rietveld refinement was performed on pXRD data with the FullProf Suite program using Thompson-Cox-Hastings pseudo-Voigt peak shapes. Parameters determined from the single crystal refinement were used as a starting point for the powder refinement. Lattice parameters and site occupancies from both powder and single crystal refinements are consistent with one another and therefore only the single crystal data will be discussed here.\\
\indent A Quantum Design Physical Property Measurement System (PPMS) Dynacool equipped with vibrating sample magnetometer (VSM) and resistivity options was used to measure the temperature and field-dependent magnetization and temperature-dependent electrical resistance of Nb$_{20.4}$Ru$_{5.7}$Ge$_{3.9}$. A standard four-probe method was used for the temperature-dependent resistance measurement taken from 300 K - 1.7 K with an applied current of 1 mA and applied magnetic fields ranging from 0 - 0.85 T. Zero-field cooled (ZFC) and field cooled (FC) magnetic susceptibility data were collected with an applied field of 10 Oe in the temperature range from 1.68 K - 3.5 K. The field-dependent magnetization was measured at various temperatures from 1.68 K - 2.2 K with a field sweep from 0 - 100 Oe. A Quantum Design PPMS Evercool II was also used to measure the heat capacity on a small crystal of Nb$_{20.4}$Ru$_{5.7}$Ge$_{3.9}$ with 0, 0.1, 0.2, 0.3, 0.4 T applied fields.

\section{Results and Discussion}

\begin{figure}[tbh!]
\includegraphics[scale = 0.38]{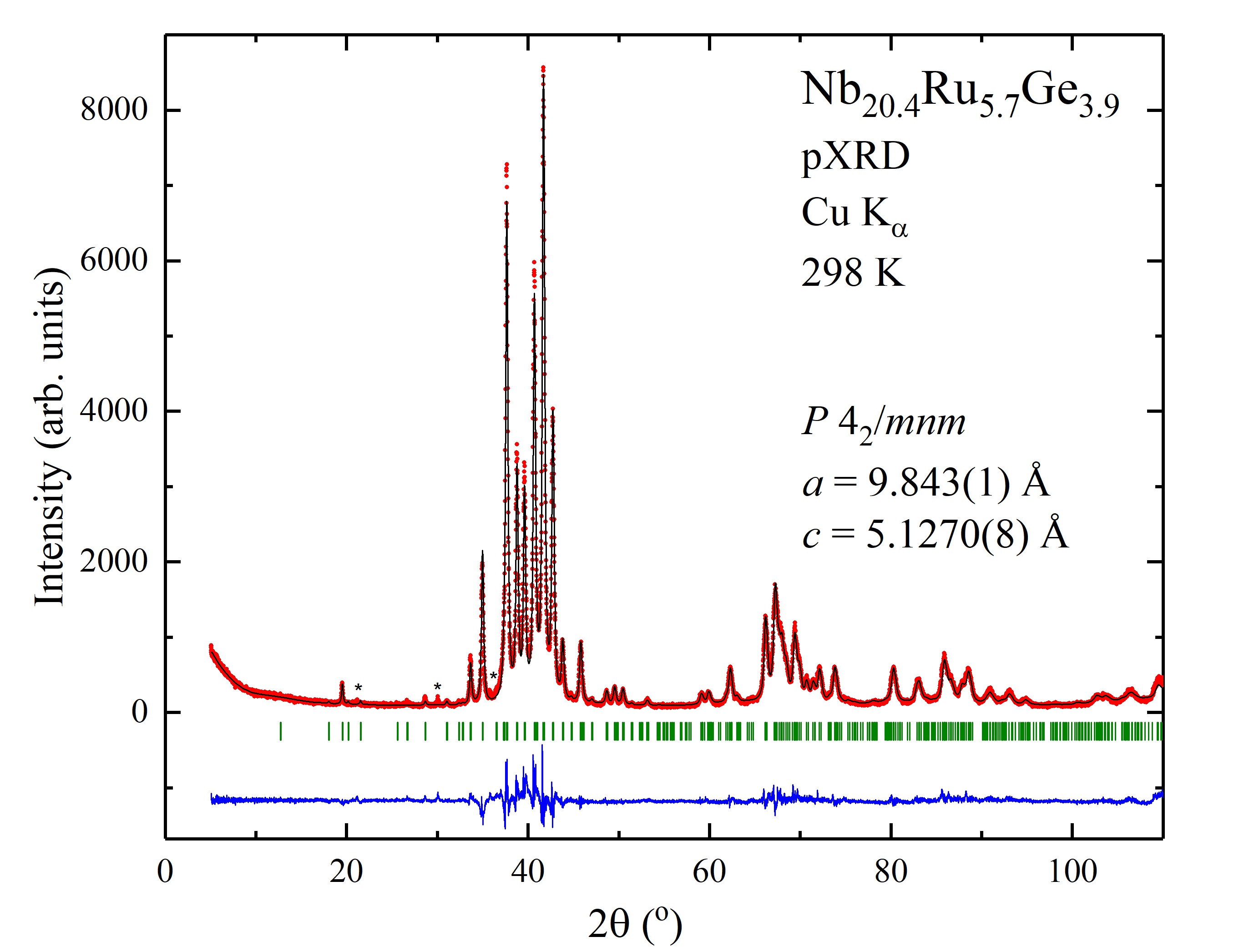}
\caption{Rietveld refinement of Nb$_{20.4}$Ru$_{5.7}$Ge$_{3.9}$ using room temperature pXRD data. The experimentally observed data is shown in red circles, the calculated pattern is shown with a black line, the green vertical marks indicate expected Bragg reflections, and the blue line at the bottom shows the difference between the observed and calculated data. Impurity peaks are marked with asterisks. Rietveld refinement results: $\chi^2$ = 3.53; $\textit{w}$R$_{p}$ = 13.6 $\%$; R$_p$ = 11.6 $\%$; R(F$^2$) = 8.36 $\%$. The cluster of strong peaks near 2$\theta$ = 40 degrees is a characteristic of $\sigma$-phases.}
\label{Fig1}
\end{figure}

\begin{figure}[t]
\includegraphics[scale = 0.55]{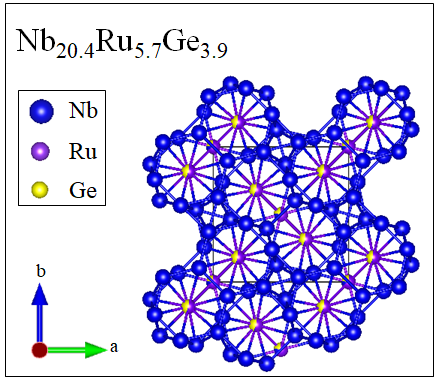}
\caption{Crystal structure of Nb$_{20.4}$Ru$_{5.7}$Ge$_{3.9}$ viewed along the $c$-direction emphasizing the topologically closest packed structure. Niobium is shown in blue, ruthenium is shown in purple, and germanium is shown in yellow.}
\label{Fig2}
\end{figure}

 Powder X-ray diffraction (pXRD) and single crystal X-ray diffraction were used to analyze the previously unreported $\sigma$-phase Nb$_{20.4}$Ru$_{5.7}$Ge$_{3.9}$, which was shown to crystallize in the CrFe structure-type ($\textit{P}$ 4$_2$/$\textit{mnm}$, No. 136) with lattice parameters $a$ = 9.843(1) $\textsc{\AA}$ and $c$ = 5.1270(8) $\textsc{\AA}$. Table \ref{Table1} shows a summary of the results from single crystal diffraction data and Table \ref{Table2} gives the atomic coordinates determined from the structure refinement. Fig. \ref{Fig1} shows the room temperature powder diffraction pattern of Nb$_{20.4}$Ru$_{5.7}$Ge$_{3.9}$ with the corresponding Rietveld fit to the data confirming the high purity of the as-melted sample. The crystal structure of this new $\sigma$-phase viewed along the $c$ - direction is presented in Fig. \ref{Fig2} showing the topologically closest packed structure, a common feature of $\sigma$-phases.\cite{FK3,FK1,FK2}
 
\begin{table}
\caption{Single crystal crystallographic data for Nb$_{20.4}$Ru$_{5.7}$Ge$_{3.9}$ at 300(2)K.}
\begin{tabular}{c c} 
\hline 
Chemical Formula & Nb$_{20.4}$Ru$_{5.7}$Ge$_{3.9}$  \\
\hline  
F.W. (g/mol); 				& 2754.63 							\\  
Space group; $Z$ 			& $P$4$_2$/$mnm$ (No. 136);1 		\\  
$a$ (${\textsc{\AA}}$)  	& 9.843(1)							\\ 
$c$ (${\textsc{\AA}}$)   	& 5.1270(8)						\\

$V$ (${\textsc{\AA}^3}$) 	& 496.7(2)						\\

\textit{hkl} ranges		 	& $-$14 $\le$ $hk$ $\le$ 14			\\
							& $-$6 $\le$ $l$ $\le$ 6			\\
Absorption Correction	 	& Numerical							\\
Extinction Coefficient		& 	0.0006(2)				\\
$\theta$ range (deg.)	 	& 2.927$-$32.044			\\
No. reflections; $R_{int}$	& 1613; 0.1114 						\\
No. independent reflections & 		472						\\
No. parameters				& 		27						\\
$R_1$; $wR_2$ (I$>$2$\delta$(I))	& 0.0650; 0.1073		\\
$R_1$; $wR_2$ (all I)		&0.1246; 0.1287					\\
Goodness of fit				& 	1.038						\\
Diffraction peak and hole (e$^-$/${\textsc{\AA}^3}$)	& 4.249; -2.985\\
\hline
\hline
\end{tabular}
\label{Table1} 
\end{table}

\begin{table}
\caption{Atomic coordinates and equivalent isotropic displacement parameters of Nb$_{20.4}$Ru$_{5.7}$Ge$_{3.9}$ at 300(2) K. U$_{eq}$ is defined as one-third of the trace of the orthogonalized U$_{ij}$ tensor (${\textsc{\AA}^2}$). $U_{eq}$ = 0.006(1)for Ru/Ge1/Nb4, 0.0049(6) for Ru/Ge2/Nb5, 0.0072(8) for Nb1, 0.0047(5) for Nb2, and 0.0043(6) for Nb3.}
\begin{tabular}{cccccc} 
\hline
\hline 
Atom & Wyck. & Occ. & $\textit{x}$ & $\textit{y}$ & $\textit{z}$ \\ 
\hline 
Ru/Ge1	& 2\textit{a} 		& 0.67(1)/0.23 	& 0				 & 0      	& 0 \\
Nb4		& 2\textit{a} 		& 0.1		 & 0				 & 0      	& 0 \\
Ru/Ge2 	& 8\textit{j} 		& 0.54(5)/0.43 & 0.1825(2) 		& 0.1825(2) &0.2506(6)\\  
Nb5 	& 8\textit{j} 		& 0.025 & 0.1825(2) 		& 0.1825(2) &0.2506(6)\\
Nb1 	& 4\textit{f} 	& 1 & 0.3961(2)  	& 0.3961(2) & 0 \\ 
Nb2		& 8\textit{i}$_2$ 	& 1 & 0.7414(2)  & 0.0670(3) 		& 0 \\ 
Nb3 	& 8\textit{i}$_1$ 		& 1 & 0.4647(2)  &  0.1280(2) &0  \\ 

\hline
\hline 
\end{tabular}
\label{Table2} 
\end{table}

\begin{figure}[b]
\includegraphics[scale = 0.35]{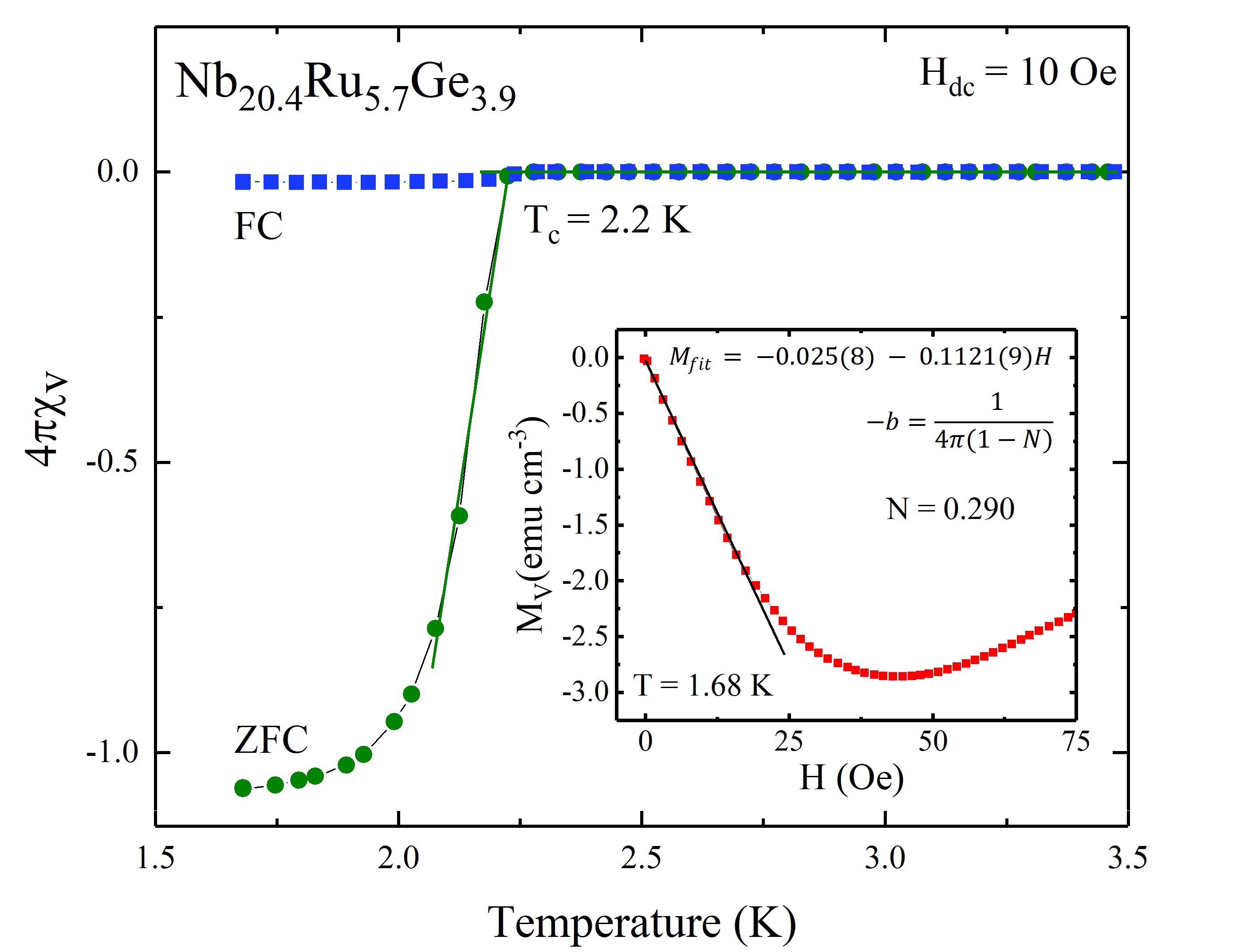}
\caption{Zero-field cooled (ZFC) and field cooled (FC) (main panel) temperature-dependent magnetic susceptibility $\chi_V$(T) measured in a 10 Oe applied magnetic field from 1.68 K - 3.5 K showing the superconducting transition at 2.2 K. The data were corrected for the demagnetization factor, N. Field-dependent magnetization (inset) measured at the lowest possible temperature, 1.68 K, to estimate the value of the demagnetization factor.}
\label{Fig3}
\end{figure}
 Binary $\sigma$-phases have the general formula $A_2B$ (normalized here to $A_{20}B_{10}$ to reflect the unit cell content). $A$ is typically an early transition element with a preference for sites with higher coordination number (CN) such as the 4$f$, 8$i_1$, and 8$j$ sites in the $\sigma$-phase structure. In contrast, the $B$ - atoms are typically more $d$-electron rich with a preference for lower CN sites like the 2$a$ and 8$i_2$ sites in the $\sigma$-phase structure. Ternary systems become more complicated, especially when a main group element like Ga, Al, or Si is included. When there are 3 elements present, it is necessary to do experiments with multiple wavelengths of radiation to quantitatively determine multiple site occupancies by diffraction. However for our purposes, since the loading composition, Nb$_{20.4}$Ru$_{5.7}$Ge$_{3.9}$, resulted in a single-phase diffraction pattern and the mass loss was $<$1$\%$ following arc-melting, the single crystal refinement was constrained to the chemical formula of the loading composition. The phase presented here, Nb$_{20.4}$Ru$_{5.7}$Ge$_{3.9}$, has the sites 4$f$, 8$i_1$, and 8$i_2$ (20 atoms total) fully occupied by Nb, with a small amount (0.4) of Nb evenly distributed across the 2$a$ and 8$j$ sites, while Ru and Ge, the ``$B$ atoms'' are mixed in different ratios on the 2$a$ and 8$j$ sites (10 atoms total). Site mixing is commonly seen in $\sigma$-phases, as previously stated. For example, Nb$_{18}$Ni$_3$Al$_9$,\cite{NbXAl} Cr$_{13.5}$Fe$_{13.5}$Si$_3$,\cite{CrFeSiSystem} Mo$_{12}$Ru$_{12}$Ta$_6$,\cite{MoRuTa} and Nb$_{18}$Mn$_6$Ga$_6$\cite{TernaryGa} all form the $\sigma$-phase but clearly have quite different combinations of elements and degrees of mixing. In addition, the examples above show that simple $A$ and $B$ element assignments are not always followed in these ternary $\sigma$-phases. Although there are numerous previously reported ternary $\sigma$-phases containing all transition metals\cite{Yaqoob2012,FeCrV_SigmaPhase} or two transition metals with Ga,\cite{TernaryGa,NbVGa,NbGaMn} Al,\cite{NbXAl,SigmaZrRh,AlTiPt} or Si\cite{CrFeSiSystem,FeVSi} as the third element, to the best of our knowledge there are no previously reported ternary $\sigma$-phases containing Ge, despite the close proximity to Al, Si, and Ga in the periodic table. In addition, the higher percentage of Nb, due to the full occupancy on the 4$f$, 8$i_1$, and 8$i_2$ sites, is similar to the binary $\sigma$-phase superconductors Nb$_{65.2}$Rh$_{34.8}$\cite{NbRh_TaRh} and Nb$_{62}$Pt$_{38}$\cite{Nb_Pt1}, and could help to explain why superconductivity is seen in this new ternary $\sigma$-phase.

 The temperature-dependent magnetic susceptibility ($\chi_V$) is shown in Fig. \ref{Fig3} for Nb$_{20.4}$Ru$_{5.7}$Ge$_{3.9}$, measured in an applied magnetic field of \textit{H} = 10 Oe. The zero field-cooled (ZFC) volume magnetic susceptibility data is only slightly less than the ideal 4$\pi\chi_V$ = $-$1 at the lowest possible temperature 1.68 K.  Both the ZFC and FC magnetic susceptibility data were corrected for a demagnetization factor (a correction for the sample shape) N equal to 0.290. The value N was calculated from the fit (M$_\texttt{fit}$) to the magnetic susceptibility vs. applied field measurements taken at 1.68 K at low fields, from 0 to 15 Oe, as shown in Fig. \ref{Fig3} (inset). Assuming linear behavior of \textit{M}$_\texttt{V}$ vs. \textit{H} in the superconducting state, the demagnetization factor can be calculated by the equation $-b = \frac{1}{4\pi(1-N)}$ , where $b$ is the slope of the linear fit and hence $\chi_V$. 

\begin{figure}[b]
\includegraphics[scale = 0.35]{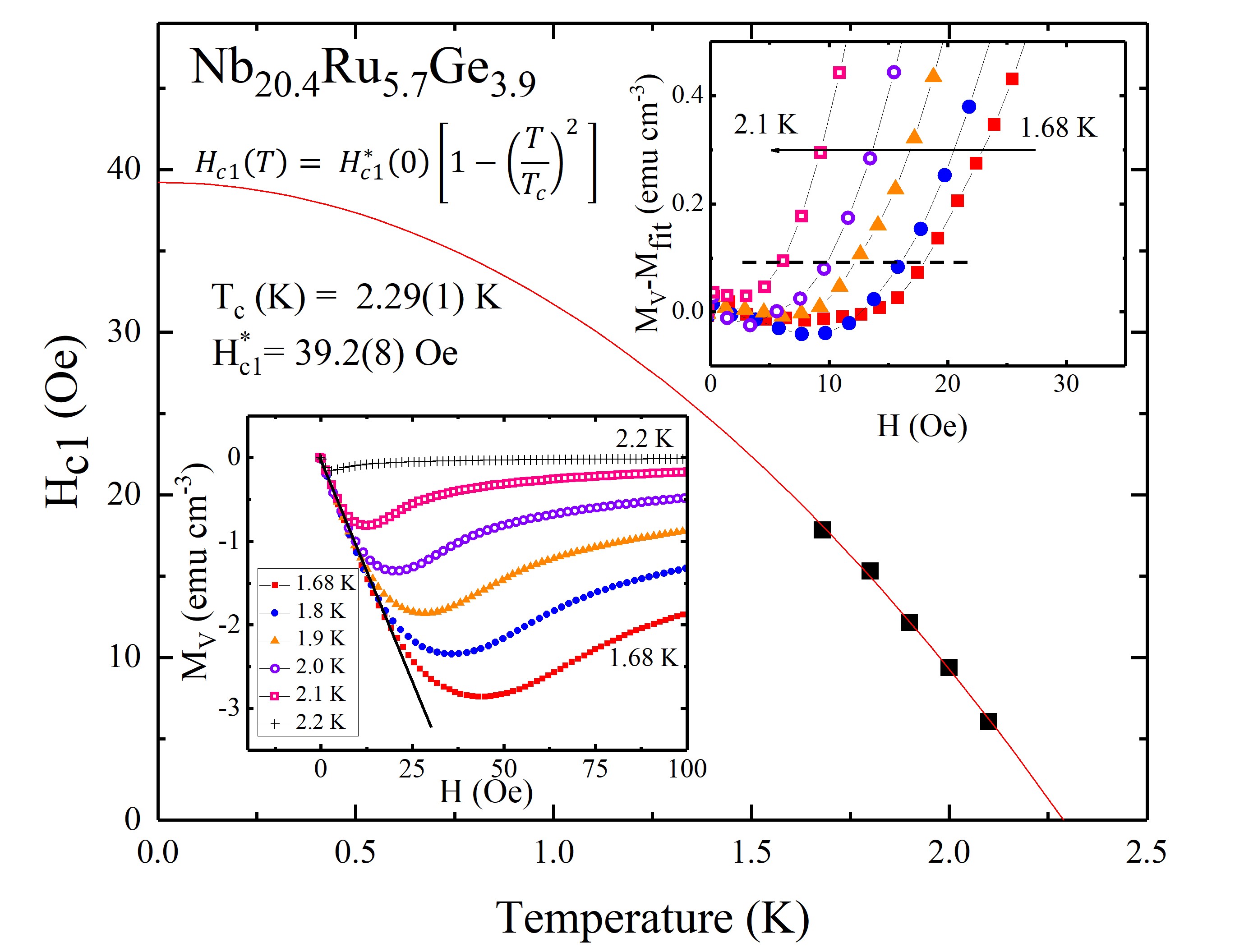}
\caption{Magnetization (\textit{M}$_\texttt{V}$) vs. applied field (\textit{H}) for the superconductor Nb$_{20.4}$Ru$_{5.7}$Ge$_{3.9}$ at temperatures between 1.68 K - 2.2 K with a field sweep from 0 - 100 Oe (lower left inset). The difference between the magnetization (\textit{M}$_\texttt{V}$) and the M$_\texttt{fit}$ at different temperatures (upper right inset). The estimation of \textit{H}$_{c1}^{*}$ from the \textit{M}$_\texttt{V}$-M$_\texttt{fit}$ plot (main panel). }
\label{Fig4}
\end{figure}

\begin{figure}[t]
\includegraphics[scale = 0.35]{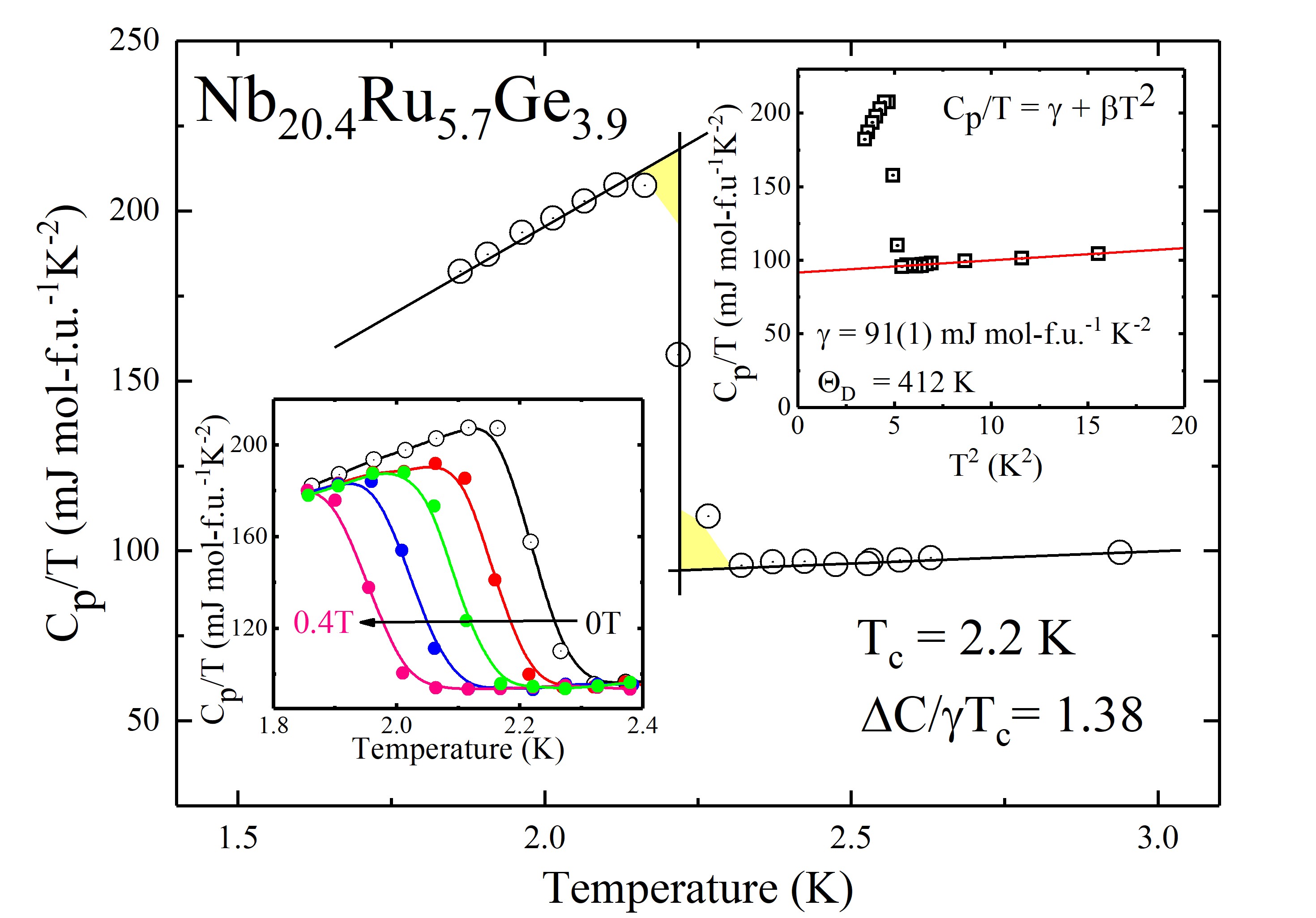}
\caption{C$_p$/T vs. T plotted from 1.8 K - 3 K measured in zero applied field where the solid black lines outline the equal area construction shown in yellow shading (main panel).  C$_p$/T vs. T for various applied magnetic fields ranging from 0 - 0.4 T increasing by 0.1 T increments (lower left inset). C$_p$/T vs T$^2$ plotted in the low temperature region fitted to a line (upper right inset). }
\label{Fig5}
\end{figure}
 Fig. \ref{Fig4} shows the characterization of the Nb$_{20.4}$Ru$_{5.7}$Ge$_{3.9}$ superconductor with field-dependent magnetization measurements. The lower left inset of Fig. \ref{Fig4} shows data taken at different temperatures ranging from 1.68 K to 2.2 K with field sweeps from 0 - 100 Oe. The difference between the magnetization (\textit{M}$_\texttt{V}$) and the M$_\texttt{fit}$ measured at 1.68 K is shown in Fig. \ref{Fig4} (upper right inset).  The fields (\textit{H}) at which there is a deviation from linearity, indicated by the dashed line, was used to construct the plot in Fig. \ref{Fig4} (main panel) plotted as a function of temperature. The \textit{H}$_{c1}$ vs. T data were fitted to equation \ref{Eq1},
\begin{equation}
H_{c1} (T) = H_{c1}^{*}(0)\left[ 1-\left( \frac{T}{T_c}\right) \right] ^2
\label{Eq1}
\end{equation}

\noindent where \textit{H}$_{c1}$(0) is the lower critical field at 0 K and \textit{T}$_c$ is the calculated critical temperature. The lower critical field, \textit{H}$_{c1}^{*}$(0), was calculated to be 39.2(8) Oe and after correcting for the demagnetization factor (N = 0.290), \textit{H}$_{c1}$(0) = 55 Oe. The calculated \textit{T}$_c$ value was 2.29(1) K, consistent with the \textit{T}$_c$ from both temperature-dependent specific heat data and resistance data, which will be discussed next.

Temperature-dependent specific heat measurements were carried out as presented in Fig. \ref{Fig5} (main panel), which plots C$_p$/T vs T in zero applied field near the transition temperature. The large anomaly in the specific heat is consistent with bulk superconductivity in Nb$_{20.4}$Ru$_{5.7}$Ge$_{3.9}$. The superconducting \textit{T}$_c$ value was determined by equal-entropy constructions of the idealized specific heat capacity jump (shown with yellow shading), which is sharp in temperature. The \textit{T}$_c$ of Nb$_{20.4}$Ru$_{5.7}$Ge$_{3.9}$ was determined to be 2.2 K, consistent with both the resistance and magnetic susceptibility data. The lower left inset of Fig. \ref{Fig5} shows the temperature dependence of the specific heat data in applied magnetic fields from 0 - 0.4 T with 0.1 T increments. The \textit{T}$_c$ is suppressed to lower temperature as the applied field is increased, as expected.

The upper right inset of Fig. \ref{Fig5} shows a plot of C$_p$/T vs T$^2$ which was fitted to equation \ref{Eq2}
\begin{equation}
\frac{C_p}{T} = \gamma + \beta T^2
\label{Eq2}
\end{equation}

\noindent where $\beta$T$^3$ is the phonon contribution and $\gamma$T is the electronic contribution to the specific heat. The Sommerfeld parameter γ was calculated to be 91(1) mJ mol-f.u.$^{-1}$K$^{-2}$ and $\beta$ was 0.831(5) mJ mol-f.u.$^{-1}$K$^{-4}$ based on the slope of the fitted line. The large gamma is caused by a large number of atoms (30) per formula unit. The Debye temperature $\Theta_\textsc{D}$ can then be calculated using $\beta$ with the following equation,
\begin{equation}
\Theta_D = \left(\frac{12\pi^4}{5\beta} n R\right)^\frac{1}{3}
\label{Eq3}
\end{equation}

\noindent where R is the gas constant 8.314 J mol$^{-1}$ K$^{-1}$ and n = 30 for Nb$_{20.4}$Ru$_{5.7}$Ge$_{3.9}$. Based on this Debye model, $\Theta_\textsc{D}$ was calculated to be 412 K. It is worth noting that the Debye temperature for elemental Ru and Ge is 600 K and 374 K, respectively.\cite{DebyeTemp} The $\Theta_\textsc{D}$ and determined \textit{T}$_c$ value can then be used to calculate the electron-phonon coupling constant $\lambda_\texttt{ep}$ from the inverted McMillan\cite{McMillan} formula as follows:
\begin{equation}
\lambda_{ep} = \frac{1.04 + \mu^*\ln\left( \frac{\Theta_D}{1.45T_c}\right)}{(1-0.62\mu^*)\ln\left( \frac{\Theta_D}{1.45 T_c}\right) - 1.04}.
\label{Eq4}
\end{equation}

\begin{figure}[t]
\includegraphics[scale = 0.35]{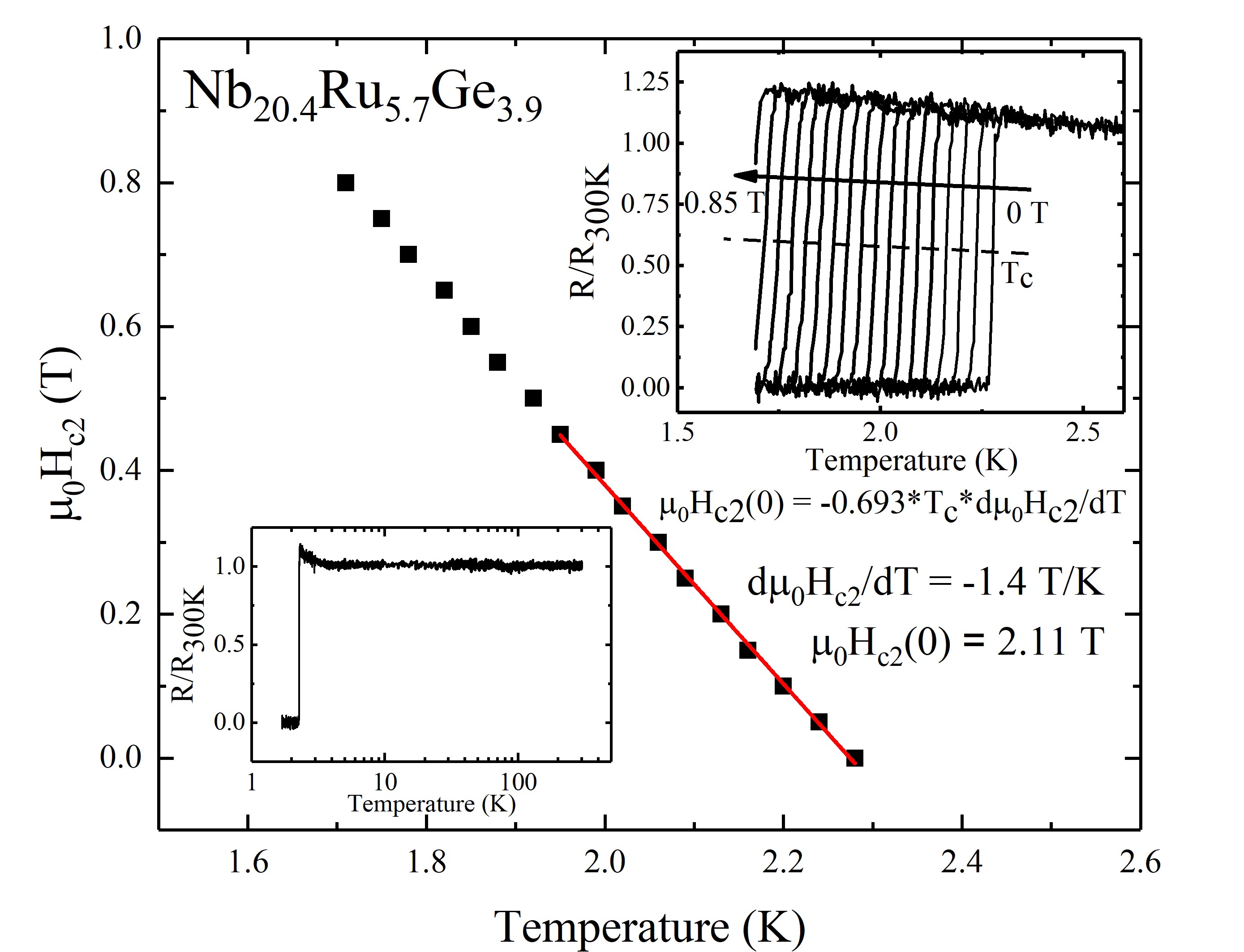}
\caption{Temperature-dependent electrical resistance normalized as R/R$_{\texttt{300K}}$ measured over the temperature range 1.7 K - 300 K with no applied field plotted on a log scale (lower left inset). Dependence of superconducting transition on applied magnetic field (upper right inset) plotted as the normalized resistance (R/R$_{\texttt{300K}}$) from 1.7 K - 2.75 K in applied magnetic fields ranging from $\mu_0$\textit{H} = 0 T to 0.85 T in steps of 0.5 T. Dashed line represents 50$\%$ of the superconducting transition. Plot of \textit{H}$_{c2}$(T) obtained from resistance data which was fitted to a line (main panel), resulting in a calculated value of $\mu_0$\textit{H}$_{c2}$(0) = 2.11 T.}
\label{Fig6}
\end{figure}

\noindent Assuming $\mu^*$= 0.13 and \textit{T}$_c$ = 2.2 K, $\lambda_\texttt{ep}$ was calculated to be 0.49 which suggests that Nb$_{20.4}$Ru$_{5.7}$Ge$_{3.9}$ is a weak coupling superconductor. The Fermi energy N(E$_\texttt{F}$) can be calculated using the equation,
\begin{equation}
N(E_F) = \frac{3\gamma}{\pi^2 k_B^2(1 + \lambda_{ep})},
\label{Eq5}
\end{equation}

\noindent where k$_\texttt{B}$ is the Boltzmann constant and $\gamma$ = 91(1) mJ mol-f.u.$^{-1}$K$^{-1}$. The estimated N(E$_\texttt{F}$) = 39 states eV$^{-1}$ per formula unit of Nb$_{20.4}$Ru$_{5.7}$Ge$_{3.9}$. The specific heat jump at transition to the superconducting state $\Delta$C/\textit{T}$_c$ was calculated to be 123 mJ mol-f.u$^{-1}$K$^{-2}$ and $\Delta$C/$\gamma$\textit{T}$_c$ = 1.38, which is close to the expected value of 1.43, confirming bulk superconductivity in Nb$_{20.4}$Ru$_{5.7}$Ge$_{3.9}$.

\begin{figure}[b]
\includegraphics[scale = 0.35]{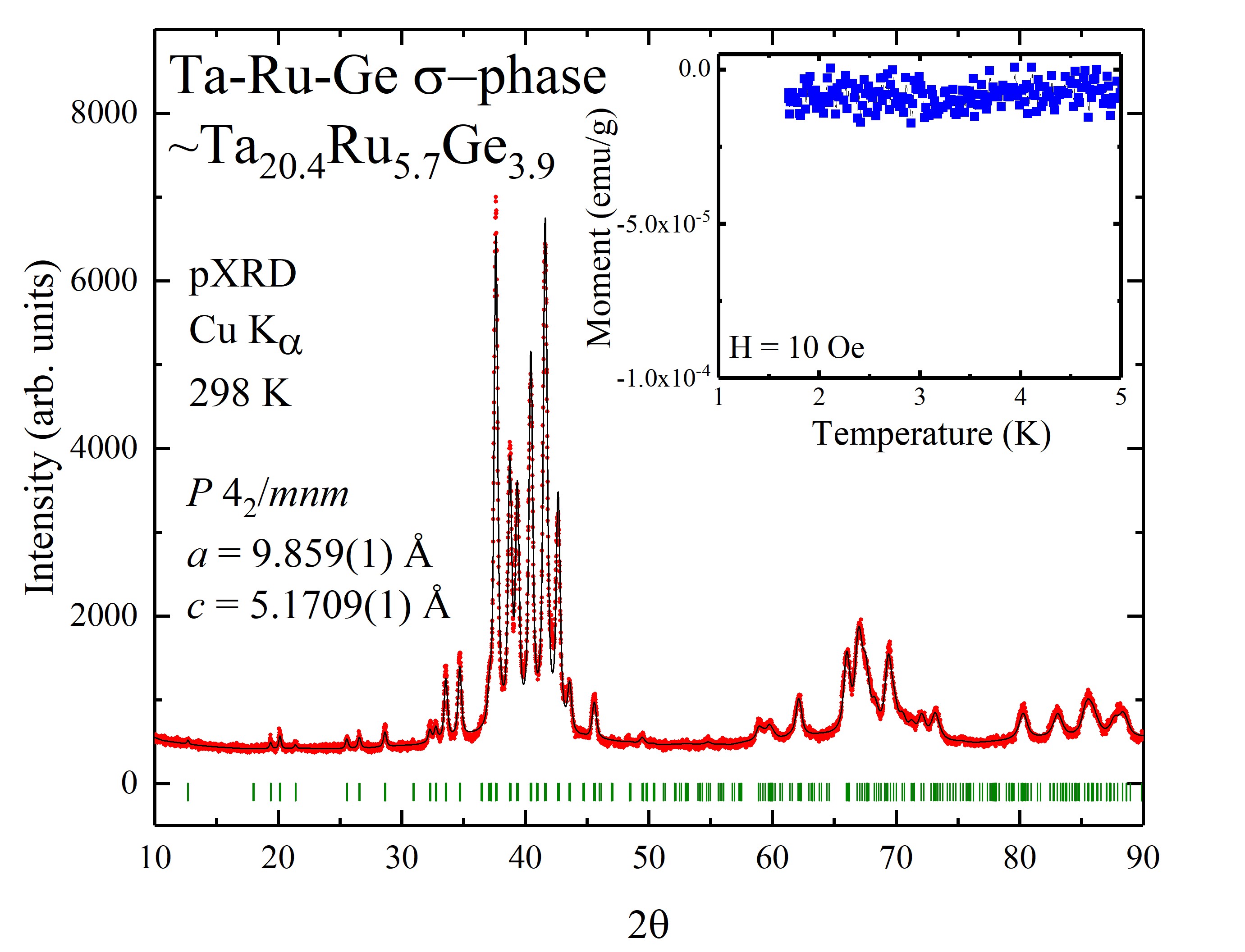}
\caption{Room-temperature pXRD pattern showing a LeBail fit of the $\sigma$-phase in the Ta-Ru-Ge system, $\sim$Ta$_{20.4}$Ru$_{5.7}$Ge$_{3.9}$ (main panel). Experimentally observed data is shown in red circles, the calculated diffraction pattern is shown with a black line, the green vertical marks indicate expected Bragg reflections. Zero-field cooled (ZFC) temperature-dependent magnetic susceptibility data (inset) measured from 1.7 K - 5 K with a \textit{H} = 10 Oe applied magnetic field, showing that the Ta-Ru-Ge $\sigma$-phase is not superconducting down to 1.7 K. }
\label{Fig7}
\end{figure}
The temperature-dependent normalized electrical resistance R/R$_{\texttt{300K}}$ for a polycrystalline, irregularly shaped sample of Nb$_{20.4}$Ru$_{5.7}$Ge$_{3.9}$ measured from  300 K - 1.7 K is shown in Fig. \ref{Fig6} (lower left inset). The resistance is relatively temperature independent in the entire temperature range from 2.3 K - 300 K, most likely due to the large extent of mixing disorder in this new material; the $\sigma$-phase in this system is a poor metal. In zero applied magnetic field, the resistance drops to zero resulting in a \textit{T}$_c$ of 2.2 K for the superconductor. There is also a slight increase in the resistance slightly above \textit{T}$_c$ where the resistance is $\sim$1.0 throughout the entire temperature range but jumps up to 1.1 at about 2.3 K.  This behavior has been previously observed in other superconductors,\cite{resistance2, resistance1, resistance3} and is most likely inherent of the material and not caused by the experimental setup. Fig. \ref{Fig6} (upper right inset) shows the dependence of the critical temperature on the applied magnetic field, where \textit{T}$_c$ was taken as 50$\%$ of the superconducting transition (dashed line). The critical temperature decreases steadily as the applied field increases from 0 - 0.85 T where the \textit{T}$_c$ is suppressed to approximately 1.71 K when $\mu_0$\textit{H} = 0.80 T. The superconducting transition remains narrow in temperature for all fields studied. The estimated \textit{T}$_c$ values from the midpoints of resistance measurements were plotted (Fig. \ref{Fig6}, main panel) and fit to a line (d$\mu_0$\textit{H}$_{c2}$/dT = $-$1.4 T/K). By using equation \ref{Eq6},
\begin{equation}
\mu_0H_{c2}(0) = -A T_c \frac{d\mu_0H_{c2}}{dT}\bigg|_{T=T_c},
\label{Eq6}
\end{equation}
\begin{table}
\caption{Superconductivity parameters of Nb$_{20.4}$Ru$_{5.7}$Ge$_{3.9}$.}
\begin{tabular}{ccc} 

\hline 
Parameter & Units & Nb$_{20.4}$Ru$_{5.7}$Ge$_{3.9}$ \\

\hline 
\textit{T}$_c$						& K		&  2.2 \\
$\mu_0$\textit{H}$_{c1}$(0) 			& mT	&  5.5\\  
$\mu_0$\textit{H}$_{c2}$(0) 			& T		&  2.11\\ 
$\mu_0$\textit{H}$_{c}$(0) 			& mT	&  60\\ 
$\xi_\texttt{GL}$ 			& $\textsc{\AA}$	&  125\\ 
$\lambda_\texttt{GL}$ 			& $\textsc{\AA}$	&  3115\\
$\kappa_\texttt{GL}$ 			& -	&  25\\
$\gamma$ 			& mJ mol-f.u.$^{-1}$K$^{-2}$	&  91\\
$\Delta$C/$\gamma$\textit{T}$_c$ 			& -	&  1.38\\
$\mu_0$\textit{H}$^\texttt{Pauli}$(0) 			& T	&  4.1\\ 
$\lambda_\texttt{ep}$ 			& -	&  0.49\\
N(E$_\texttt{F}$)			& states eV$^{-1}$ per f.u.	&  39\\
$\Theta_\textsc{D}$				& K				& 412\\

\hline
\hline 
\end{tabular}
\label{Table3} 
\end{table}
\begin{figure}[b]
\includegraphics[scale = 0.35]{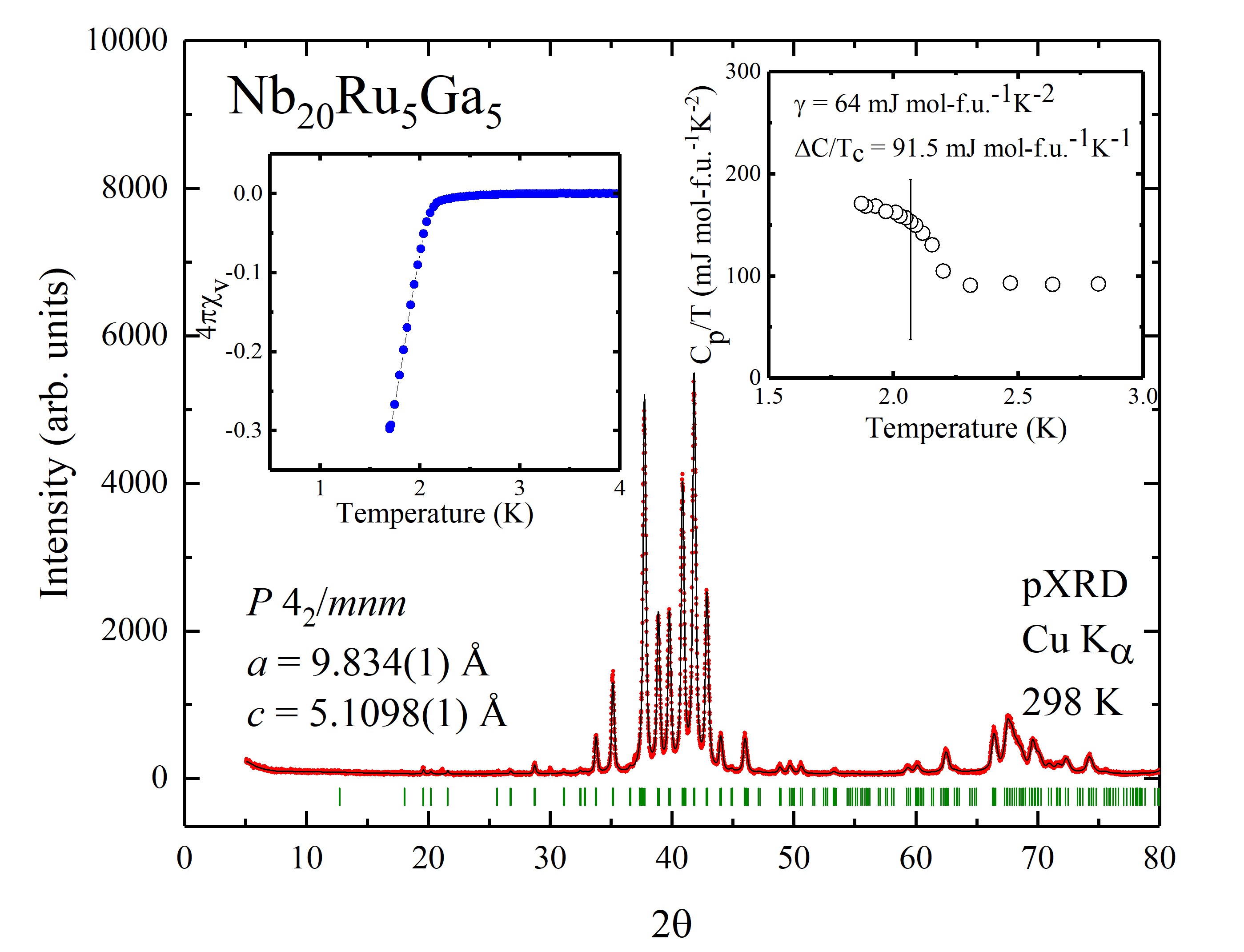}
\caption{Room-temperature pXRD pattern showing a LeBail fit of the $\sigma$-phase Nb$_{20}$Ru$_5$Ga$_5$ (main panel). Experimentally observed data is shown in red circles, the calculated diffraction pattern is shown with a black line, the green vertical marks indicate expected Bragg reflections for space group $\textit{P}$4$_2$/$\textit{mnm}$. Preliminary temperature-dependent specific heat data showing an anomaly in the specific heat at $\sim$2 K (upper right inset). Zero-field cooled (ZFC) temperature-dependent magnetic susceptibility data (upper left inset) showing an incomplete superconducting transition at $\sim$2K.}
\label{Fig8}
\end{figure}

\noindent with a \textit{T}$_c$ = 2.2 K, and A is 0.69 for the dirty limit or 0.73 for the clean limit,\cite{cleanlimit} $\mu_0$\textit{H}$_{c2}$(0) was calculated to be 2.11 T and 2.25 T for the dirty and clean limit of Nb$_{20.4}$Ru$_{5.7}$Ge$_{3.9}$, respectively. Both values are lower than the Pauli limit $\mu_0$\textit{H}$^\texttt{Pauli}$ = 1.85*\textit{T}$_c$ = 4.1 T. The Ginzburg-Landau superconducting coherence length $\xi_\texttt{GL}$ was estimated to be 125 $\textsc{\AA}$ from the equation 
\begin{equation}
H_{c2}(0) = \frac{\Phi_0}{2\pi\xi_{GL}^2}
\label{Eq7}
\end{equation}

\noindent where $\Phi_0$ = h/2e and $\mu_0$\textit{H}$_{c2}$(0) = 2.11 T. The lower critical field, \textit{H}$_{c1}$(0) = 55 Oe, was used with $\xi_\texttt{GL}$ = 125 $\textsc{\AA}$ to estimate the superconducting penetration depth $\lambda_\texttt{GL}$ to be 3115 $\textsc{\AA}$ using the formula, 
\begin{equation}
H_{c1} = \frac{\Phi_0}{4\pi\lambda_{GL}^2}\ln\frac{\lambda_{GL}}{\xi_{GL}}
\label{Eq8}
\end{equation}

The ratio of the calculated values of $\lambda_\texttt{GL}$ and $\xi_\texttt{GL}$ will give the value $\kappa_\texttt{GL}$= 25 [$\kappa_\texttt{GL}$ =$\lambda_\texttt{GL}$/ $\xi_\texttt{GL}$], confirming type-II superconductivity in Nb$_{20.4}$Ru$_{5.7}$Ge$_{3.9}$. In addition, the $\kappa_\texttt{GL}$ value can be used in the equation
\begin{equation}
H_{c1}H_{c2} = H_c^2 \ln\kappa_{GL}
\label{Eq9}
\end{equation}
to calculate the thermodynamic critical field $\mu_0$\textit{H}$_{c}$ = 60 mT. A summary of all superconducting parameters is given in Table \ref{Table3}. A Ta-Ru-Ge $\sigma$-phase, with an approximate composition of Ta$_{20.4}$Ru$_{5.7}$Ge$_{3.9}$, was synthesized as shown in Fig. \ref{Fig7} (main panel). The Ta-variant $\sigma$-phase was tested for superconductivity but did not display superconductivity above 1.7 K (Fig. \ref{Fig7}, inset). This suggests that electron count does not strictly govern the superconducting transition temperature seen in $\sigma$-phases.  Finally, in Fig. \ref{Fig8}, we show preliminary powder X-ray diffraction (main panel), magnetic susceptibility (left inset), and specific heat data (right inset) for the $\sigma$-phase in the Nb-Ru-Ga system, whose composition is Nb$_{20}$Ru$_5$Ga$_5$, suggesting the presence of Nb is crucial for the superconductivity observed. The superconducting \textit{T}$_c$ appears to be 2.1 K. Lower temperature measurements are required to fully characterize the superconducting transition.

\section{Conclusions}
We report the new ternary $\sigma$-phase superconductor Nb$_{20.4}$Ru$_{5.7}$Ge$_{3.9}$ which is shown to adopt the CrFe structure-type ($\textit{P}$ 4$_2$/$\textit{mnm}$, No. 136). Single crystal diffraction studies showed that Nb fully occupies the 4$f$, 8$i_1$, and 8$i_2$ sites, a small amount of Nb was evenly distributed on the 8$j$ and 2$a$ sites, while there is Ru/Ge mixing on the remaining 2$a$ and 8$j$ sites. Temperature-dependent magnetic susceptibility, resistance, and specific heat measurements show that the material enters the superconducting state below a critical temperature of 2.2 K and that the superconductivity is an intrinsic property of the material. Based on the calculated superconducting parameters, Nb$_{20.4}$Ru$_{5.7}$Ge$_{3.9}$ is a weak coupling type II BCS superconductor. The $\sigma$-phase with approximate composition Ta$_{20.4}$Ru$_{5.7}$Ge$_{3.9}$ was synthesized but does not display superconductivity above 1.7 K. We also present preliminary specific heat, pXRD, and magnetic susceptibility results for the potential 2 K superconductor Nb$_{20}$Ru$_5$Ga$_5$. Measurements down to 0.4 K would be of interest for future work to fully characterize this possible superconductor. Thus, though the \textit{T}$_c$ values are low, our results combined with those in the literature for binary alloys, suggest that $\sigma$-phase alloys appear to be favorable hosts for superconductivity. 
\section*{Acknowledgements}
The materials synthesis was supported by the Department of Energy, Division of Basic Energy Sciences, Grant DE-FG02-98ER45706, and the property characterization was supported by the Gordon and Betty Moore Foundation EPiQS initiative, Grant GBMF-4412. The work at LSU was supported by start-up funding through the LSU-College of Science. The research in Poland was supported by the National Science Centre, Grant UMO-2016/22/M/ST5/00435.  

\section*{Author Correspondence}
\begin{flushleft}
$^*$E.M.C (carnicom@princeton.edu)\\
$^*$R.J.C. (rcava@exchange.princeton.edu)
\end{flushleft}

\bibliographystyle{apsrev4-1}
\bibliography{MyBib}

\end{document}